\documentclass[12pt]{article}
\usepackage{epsfig}
\usepackage{amssymb}
\usepackage{amsfonts}
\usepackage{subfigure}
\usepackage{color}
\def\beq{\begin{equation}}
\def\eeq{\end{equation}}
\def\bea{\begin{eqnarray}}
\def\eea{\end{eqnarray}}
\def\ksl{\hbox{\hbox{${k}$}}\kern-1.9mm{\hbox{${/}$}}}

\newcommand{\text}{\rm}

\def\lsim{\raise0.3ex\hbox{$\;<$\kern-0.75em\raise-1.1ex\hbox{$\sim\;$}}} 

\def\gsim{\raise0.3ex\hbox{$\;>$\kern-0.75em\raise-1.1ex\hbox{$\sim\;$}}}

\begin{document}

\begin{center} 
{\bf \large  Stability constraints of the scalar potential in extensions \\of the Standard Model with TeV scale right handed neutrinos}
\footnote{Presented at \emph{NOW 2014}, 7-14 September 2014, Conca Specchiulla (Otranto-Italy)}

\vspace{0.2cm}
{\bf \Large } 

\vspace{1.5cm}
{\bf Claudio Corian\`o, Luigi Delle Rose and Carlo Marzo}
\vspace{1cm}

{\it Dipartimento di Matematica e Fisica "Ennio De Giorgi", 
Universit\`{a} del Salento and \\ INFN-Lecce, Via Arnesano, 73100 Lecce, Italy\footnote{claudio.coriano@le.infn.it, luigi.dellerose@le.infn.it, carlo.marzo@le.infn.it}}\\

\vspace{.5cm}
\begin{abstract} 
We investigate the stability of the scalar potential in a minimal $U(1)'$ extension of the Standard Model augmented by an extra complex scalar and three right handed neutrinos. We consider a type-I seesaw scenario for the generation of the small neutrino masses. Moreover, we focus our analysis on a class of models potentially accessible at the LHC experiment, in which the vacuum expectation value of the extra scalar is chosen in the TeV range. We show that the requirements of stability of the potential within the perturbative evolution under the renormalization group impose significant constraints on the Yukawa couplings of the right handed neutrinos.
\end{abstract}
\end{center}

\newpage

\section{Introduction}
The discovery of the Higgs boson at the LHC experiment has revived the interest in the issue of the electroweak vacuum stability of the Standard Model (SM) under the renormalization group (RG) evolution. This topic has been extensively addressed in the past \cite{Cabibbo:1979ay, Lindner:1985uk, Lindner:1988ww, Ford:1992mv}, also in connection with the requirement of the absence of a Landau pole. Under the \emph{big desert} assumption, namely the nonappearance of new significant physics effects up to the unification scale and beyond, the condition of vacuum stability of the SM scalar potential  under the perturbative RG evolution has set lower and upper theoretical bounds on the Higgs mass and on the size of the Yukawa coupling of the top quark \cite{Bezrukov:2012sa,Buttazzo:2013uya}. In these analyses, the magnitude and the sign of the different contributions to the $\beta_\lambda$ function of the Higgs quartic coupling $\lambda$ is of a paramount importance. Indeed, the 
effect of the fermions, in particular of the top quark, on the RG evolution is to drive $\lambda$ towards negative values, spoiling the vacuum stability already at $10^{9-10}$ GeV, and confining the SM to a metastable phase. In this respect, a precise measurement of the value of top mass and the solution of the ambiguities in the use of different renormalization schemes for its evaluation are mandatory \cite{Masina:2012tz}. \\
This scenario can obviously be ameliorated in various SM extensions, starting from the simplest ones. In \cite{Coriano:2014mpa} we have discussed the vacuum stability issue and the related constraints in a rather minimal extension of the SM, in which the gauge group is enlarged by an extra $U(1)'$ symmetry and an extra scalar whose vacuum expectation value $v'$ lies in the TeV range. This is a region of the parameter space which could be explored at the LHC in the next few years and which has been extensively analyzed in \cite{Basso:2010jm,Basso:2010yz,Basso:2011na,Basso:2013vla}. In this class of models, the cancellation of the $U(1)'$ gauge and the gravitational anomalies naturally predicts three right handed neutrinos $\nu_R$, singlets under the SM gauge group. For the generation on the small SM neutrino masses we have required a type-I seesaw mechanism. We have shown that the requirement of vacuum stability constraints significantly the mass $m_{\nu_h}$ of the extra right handed neutrinos, the mass of the heavy Higgs $m_{h_2}$ and the mixing angle $\theta$ in the scalar sector. Here we review some of the major results and we refer to \cite{Coriano:2014mpa} for more details.
\section{The vacuum stability analysis}
We constrain the interaction of the matter with the $Z'$ gauge boson by requiring the cancellation of the gauge and gravitational anomalies. This fixes all but two of the $U(1)'$ charges. The free ones can be used to differentiate among several $U(1)'$ models included in this extension. In the following we restrict our results to the $U(1)_{B-L}$, where $B$ and $L$ stand respectively for the baryon and the lepton numbers. For definiteness, we have chosen the new gauge coupling $g'=0.1$ and we have consistently allowed for a kinetic mixing. 

The scalar potential is characterized by the usual Higgs doublet $H$ and a by new complex scalar $\phi$, which is charged only under the new abelian gauge group $U(1)'$
\begin{eqnarray}
V(H,\phi) &=& m_1^2 H^\dag H + m_2^2 \phi^\dag \phi + \lambda_1 (H^\dag H)^2 \nonumber \\
&+&  \lambda_2  (\phi^\dag \phi)^2 + \lambda_3  (H^\dag H) ( \phi^\dag \phi) \,,
\end{eqnarray}
and whose quartic couplings are constrained by the following conditions (vacuum stability)
\begin{equation}
\lambda_1 > 0\,, \quad \lambda_2 >0 \,, \quad 4 \lambda_1 \lambda_2 - \lambda_3^2 > 0 \,.
\end{equation}

The Yukawa lagrangian is given by the usual SM one, augmented by two additional terms, introduced for the generation of the neutrino masses by a type-I seesaw mechanism
\begin{equation}
\mathcal L_{yuk} =  \mathcal L_{SM\,yuk} - Y_\nu \, L \cdot H \nu_R^c - Y_N \, \phi \, \nu_R \nu_R + h.c. \,.
\end{equation}
Having selected a value of $v'$ in the TeV range, the Yukawa $Y_\nu$ must be $\lesssim 10^{-6}$ in order to reproduce the light neutrino masses. Such a value, therefore, is too small to affect the RG evolution. On the other hand, the Yukawa $Y_N$, which is responsible for the generation of a Majorana mass term, could be $\sim O(1)$, playing a significant role in the requirement of stability of the vacuum.

All the couplings in the lagrangian evolve with RG equations that are too lengthy to be reported here. We just want to mention that our results are based on a re-analysis of the $\beta$ functions of the quartic couplings of the scalar potential. Indeed, $\beta_{\lambda_2}$ differs from the expression given in some of the previous literature in regards to the coefficient of the $Y_N^4$ contribution (see for instance \cite{Datta:2013mta}). This was also pointed out in \cite{Lyonnet:2013dna,Basso:2013vla}. This change has a considerable effect on the running of $\lambda_2$ which, for certain values of $Y_N$, may become negative, destabilizing the scalar potential.

\begin{figure}[t]
\centering
\includegraphics[scale=0.45]{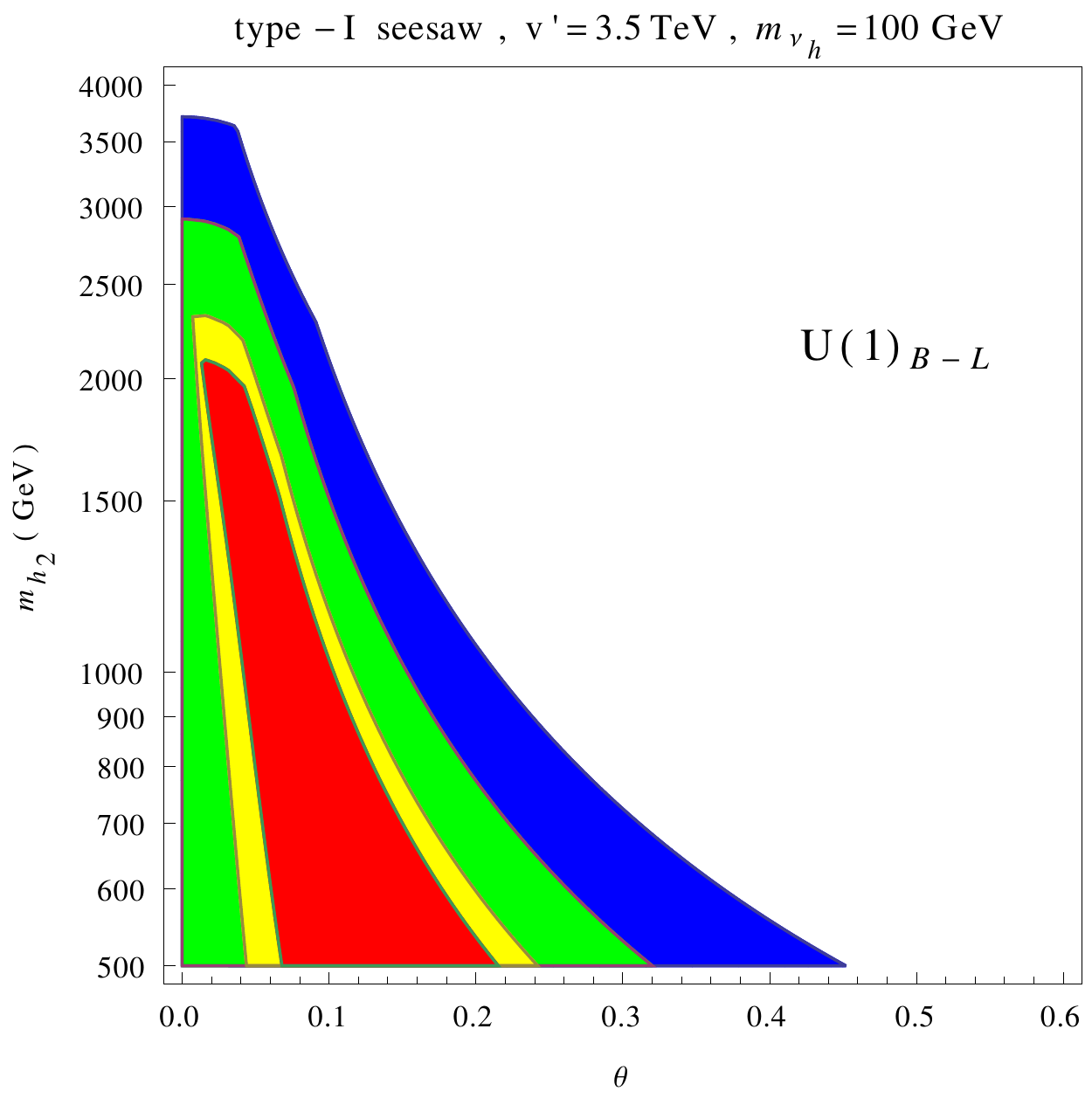}
\caption{Regions in the $(m_{h_2},\theta)$ parameter space in which the stability conditions are preserved up to $10^5$ (blue), $10^{9}$ (green), $10^{15}$ (yellow) and $10^{19}$ GeV (red) with $v'=3.5$ TeV and $m_{\nu_h} = 100$ GeV. \label{Fig.typeImh2th}}
\end{figure}
\begin{figure}[t]
\centering
\includegraphics[scale=0.6]{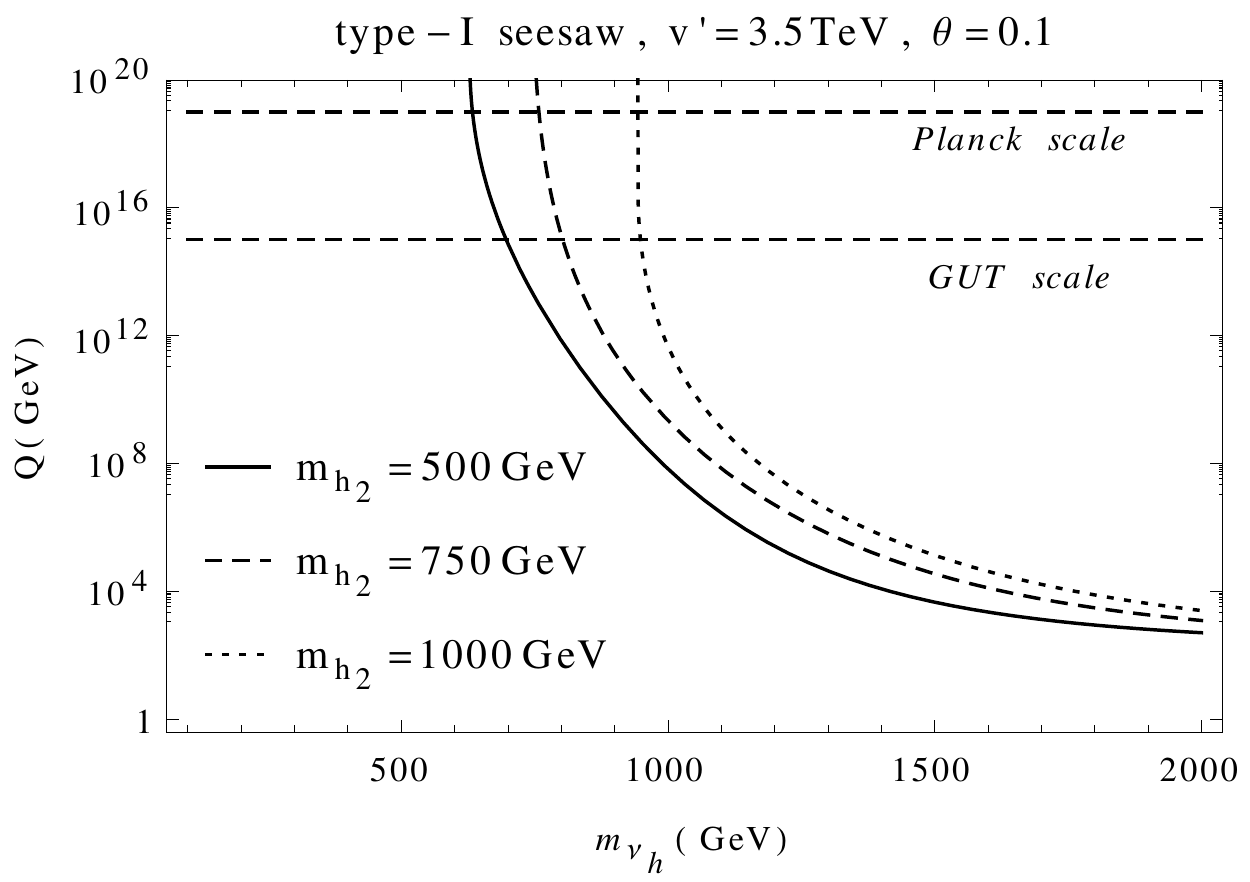} 
\caption{The maximum scale up to which the stability conditions are fulfilled as a function of the heavy neutrino mass $m_{\nu_h}$, with $v'=3.5$ TeV and $\theta = 0.1$. \label{Fig.typeIL2}}
\end{figure}
In Fig.\ref{Fig.typeImh2th} we show the stability regions in the $(m_{h_2}, \theta)$ space, where $m_{h_2}$ is the mass of the heavy scalar in the basis of the mass eigenstates, and $\theta$ is the mixing angle. The lighter $h_1$ state is identified with the observed Higgs and its mass fixed at 125 GeV. Concerning the heavy neutrinos, for the sake of simplicity, we assume their masses degenerate in flavour space. In Fig.\ref{Fig.typeImh2th} we have used $m_{\nu_h} = 100$ GeV. For $m_{h_2} \lesssim 2$ TeV, the stability regions correspond to an interval defined by small values of the mixing angle $\theta$. The same interval overlaps with the bound found in \cite{Dawson:2009yx} through the analysis of the $S,T,U$ parameters. It is very interesting that the smallness of $\theta$ can be deduced also from the stability issue, beside the electroweak precision data. 

In Fig.\ref{Fig.typeIL2} we show the maximum scale, up to which the stability is maintained, as a function of $m_{\nu_h}$ for different values of the heavy Higgs mass. It is easy to see that, in order to achieve the vacuum stability up to the GUT or Planck scales with a $v'=3.5$ TeV, the heavy neutrino mass cannot be larger than $600-1000$ GeV. Notice also that the allowed region for $m_{\nu_h}$ increases for bigger values of $m_{h_2}$. This is due to the fact that a heavier $h_2$ implies larger values of the scalar quartic couplings at the electroweak scale, which compensate the decreasing effect of a heavier $\nu_h$.

All the previous results have been obtained using the minimum allowed value by the LEP-II constraints \cite{Cacciapaglia:2006pk} for the vacuum expectation value of the extra SM singlet scalar, namely $v'=3.5$ TeV. Concerning the stability regions in the $(m_{h_2}, \theta)$ parameter space, a heavier $v'$ extends the maximum allowed values of the heavy scalar mass and has an almost negligible effect on the small $\theta$ region. On the other hand, the stability bounds on the heavy neutrino masses obviously scale with $v'$. This is due to the fact that the RG evolution of the quartic couplings depend explicitly on the dimensionless $Y_N$. Indeed, in the type-I seesaw scenario this parameter is related to the heavy neutrino mass by the relation $m_{\nu_h} = \sqrt{2} Y_N v'$.
\section{Conclusions}
We have reviewed some of the results presented in \cite{Coriano:2014mpa} in which a numerical study of the RG evolution for a generic abelian extension of the SM has been considered. In this class of models, an additional complex scalar, needed to spontaneously break the $U(1)'$ symmetry, and three heavy right handed neutrinos are naturally included. We have also implemented a type-I seesaw mechanism for the generation of the small masses of SM neutrinos. Our work has been focused on an appealing scenario, potentially accessible at the LHC, in which the vacuum expectation value of the extra scalar lays in the TeV range. We have shown that within this scenario, for a heavy scalar at the TeV scale and three right handed neutrinos of mass up to 1 TeV, all the bounds implied by the vacuum stability requirement are satisfied. On the other hand, larger mass values of the right handed neutrinos prevent to extend the validity of these models up to the GUT or Planck scales.

\end{document}